\documentclass[twocolumn,showpacs,preprintnumbers,amsmath,amssymb,prl]{revtex4-1}
\pdfoutput=1
\usepackage[T1]{fontenc}
\usepackage[utf8]{inputenc}
\usepackage{textcomp}
\usepackage{graphicx}
\usepackage{braket}
\begin{document}
\title{Spin-resolved electronic response to the phase transition in MoTe$_2$}

\author{Andrew P. Weber$^{1,2,3}$}

\email{andrew.weber@dipc.org}

\author{Philipp Rü{\ss}mann$^{4}$}

\author{Nan Xu$^{1,2}$}

\author{Stefan Muff$^{1,2}$}
\author{Mauro Fanciulli$^{1,2}$}

\author{Arnaud Magrez$^{1}$}
\author{Philippe Bugnon$^{1}$}
\author{Helmuth Berger$^{1}$}

\author{Nicholas C. Plumb$^{2}$}
\author{Ming Shi$^{2}$}

\author{Stefan Blügel$^{4}$}
\author{Phivos Mavropoulos$^{4,5}$}
\author{J. Hugo Dil$^{1,2}$}

\affiliation{$^{1}$Institute of Physics, \'{E}cole Polytechnique F\'{e}d\'{e}rale de Lausanne, CH-1015, Lausanne, Switzerland}
\affiliation{$^{2}$Swiss Light Source, Paul Scherrer Institute, CH-5232 Villigen, Switzerland}
\affiliation{$^{3}$Donostia International Physics Center, 20018 Donostia, Gipuzkoa, Spain}
\affiliation{$^{4}$Peter Grünberg Institut and Institute for Advanced Simulation, Forschungszentrum Jülich and JARA, 52425 Jülich, Germany}
\affiliation{$^{5}$Department of Physics, National and Kapodistrian University of Athens, 15784 Zografou, Greece}

\date{\today}
\begin{abstract}
The semimetal MoTe$_2$ is studied by spin- and angle- resolved photoemission spectroscopy to probe the detailed electronic structure underlying its broad range of response behavior. A novel spin-texture is uncovered in the bulk Fermi surface of the non-centrosymmetric structural phase that is consistent with first-principles calculations. The spin-texture is three-dimensional, both in terms of momentum dependence and spin-orientation, and is not completely suppressed above the centrosymmetry-breaking transition temperature. Two types of surface Fermi arc are found to persist well above the transition temperature. The appearance of a large Fermi arc depends strongly on thermal history, and the electron quasiparticle lifetimes are greatly enhanced in the initial cooling. The results indicate that polar instability with strong electron-lattice interactions exists near the surface when the bulk is largely in a centrosymmetric phase.

\end{abstract}


\maketitle

MoTe$_2$ exhibits a range of phenomena intersecting the physics of polar lattice transitions, topological phases of matter, and novel magnetoelectric properties. The centrosymmetric 1T' crystal undergoes a first order transition into the noncentrosymmetric T$_d$ structural phase upon cooling through $T_{S}\approx250$ K, with volume fractions of both phases appearing within the $200$-$300$ K range \cite{Clarke1978,Chen2016nano,Sakai2016,Yan2017a,Heikes2018}. Such transitions are very rare in metals and allow for control over the appearance of Weyl semimetal phases of matter (WSPs) and momentum dependent spin-polarization (spin-texture) that would be desirable for spintronic applications \cite{He2018a}. Superconductivity proposed to be topologically non-trivial has been observed \cite{Qi2016,Luo2016,Takahashi2017,Guguchia2017}. Like to WTe$_2$ \cite{Soluyanov2015,Ali2014}, T$_d$-MoTe$_2$ is a type-II Weyl semimetal candidate \cite{Sun2015a,Zwang2016} and exhibits extreme transverse magnetoresistence (XMR) with turn-on behavior \cite{Pei2017,ChenDec2016,Thirupathaiah2017}. Simultaneous tuning of electronic properties and the structural transition temperature and the breadth of the mixed-phase region is realized as a function of doping \cite{Sakai2016} and pressure/strain \cite{Takahashi2017,Yang2017elast,Heikes2018}. The sizes and shapes of the bulk electron Fermi pockets (EPs) and hole Fermi pockets (HPs) are important to the electronic basis for the properties of (Mo/W)Te$_2$ \cite{Ivo2014}, but there is growing recognition that responses of electronic state vectors, described in terms of their spin and/or orbital projections, play a central role \cite{Jiang2015,ChenDec2016,Yang2017elast,Pei2018,Titus2016}.

The WSP is predicted to be sensitive to the lattice parameters \cite{Sun2015a,Tamai2016,Zwang2016} and cannot exist in the centrosymmetric 1T' crystal, wherein all of the bulk bands must be spin-degenerate. However, the electronic structure of 1T'-MoTe$_2$ ($T>T_S$) observed in photoemission spectroscopy appears much the same as that of T$_d$-MoTe$_2$ ($T<<T_S$) \cite{Crepaldi2017}, although the decay of photoexcited states is clearly affected (likely due to loss of the WSP) \cite{Crepaldi2017a}. Different reports on T$_d$-MoTe$_2$ favor the case of zero (trivial semimetal) \cite{Rhodes2017}, four \cite{Zwang2016,Xu2016,Ruessmann2018,Berger2018}, or eight \cite{Tamai2016,Jiang2017,Liang2016,Deng2016,Huang2016,Sakano2017} Weyl points (WPs) in the Brillouin zone (BZ) at locations ranging from approximately 5 \cite{Huang2016} to 55 meV \cite{Ruessmann2018} above the Fermi energy $E_F$. The WPs impose subtle constraints on surface Fermi arc dispersions in (Mo/W)Te$_2$ systems \cite{Bruno2016,Tamai2016,Belopolski2016}, which have been taken as experimental signatures of the WSP \cite{Tamai2016,Deng2016,Huang2016,Liang2016,Xu2016,Sakano2017,Jiang2017,Wang2016,Wu2016,Bruno2016,Sanchez-Barriga2016,Belopolski2016,Crepaldi2017,Berger2018}. Two types of Fermi arc are present \cite{Tamai2016,Crepaldi2017,Ruessmann2018}. Small arcs are buried within the HPs and a large arc appears in the gap between the HPs and EPs. The large arc persists in 1T'-MoTe$_2$ \cite{Crepaldi2017} and in the absence of WPs in WTe$_2$ \cite{Ruessmann2018,Bruno2016}, reinforcing the fact that Fermi arcs provide insufficient (although necessary) evidence of a WSP \cite{Xu2017prl}. Quasiparticle scattering of the Fermi arcs \textit{is} strongly affected by the structural transition \cite{Berger2018}, however, this scattering occurs as a function of spin-texture and nesting conditions rather than being directly related to the WSP \cite{Ruessmann2018}. 

\begin{figure}
\includegraphics[width=0.5\textwidth]{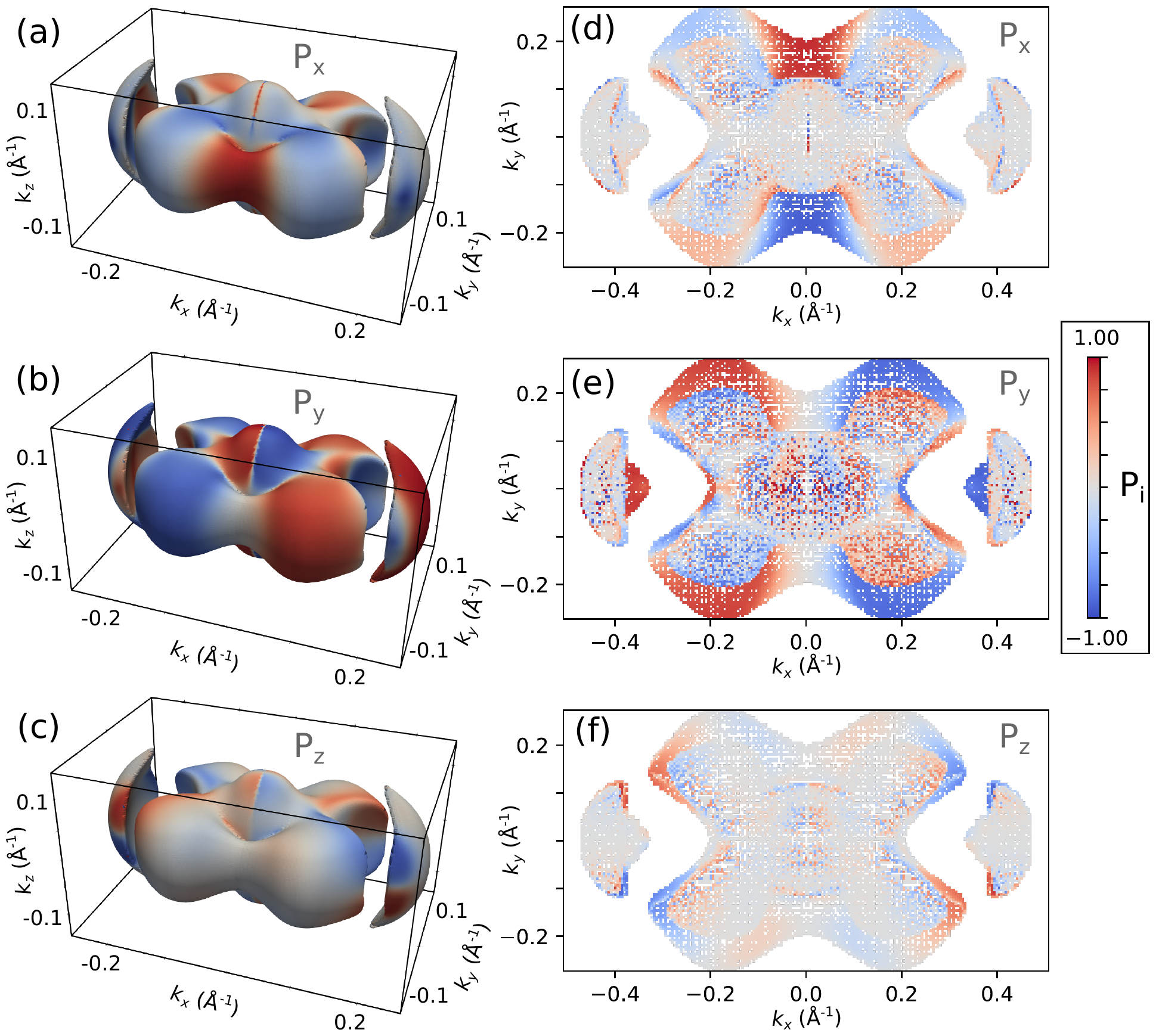} 
\caption{(Color online) Results of first-principles calculations for T$_d$-MoTe$_2$. False-color maps of (a) $P_x$, (b) $P_y$, and (c) $P_z$ on the full bulk Fermi surface and the corresponding average of (d) $P_x$, (e) $P_y$, and (f) $P_z$ over the interval $-\pi<k_z<0$ projected into the $k_x, k_y$ plane.}
\end{figure}

\begin{figure}
\includegraphics[width=0.5\textwidth]{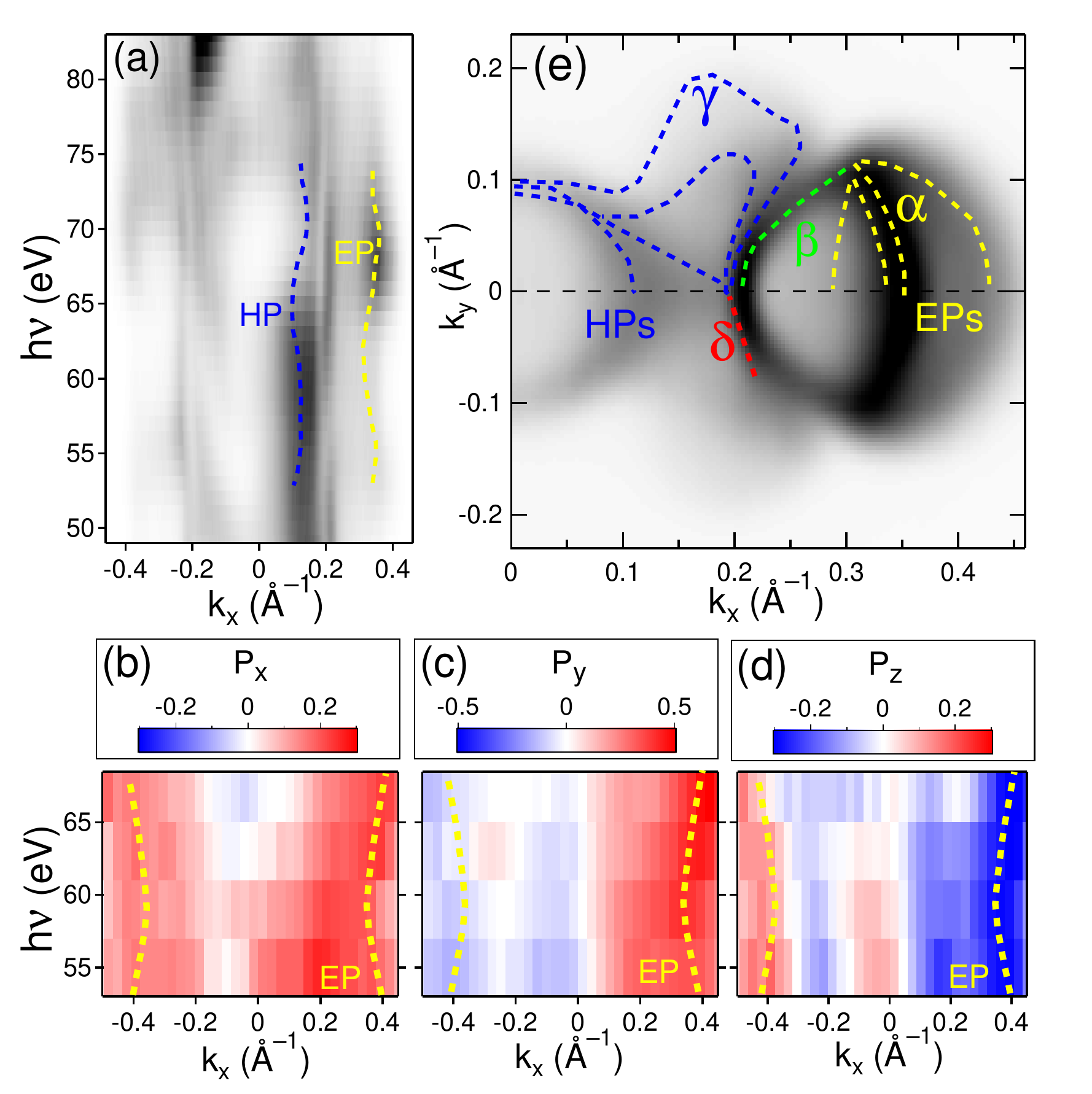}
\caption{(Color online) Photoemission data collected for T$_d$-MoTe$_2$ at T = 30 K. Photon energy dependence at $E_F$ of (a) ARPES intensity along $\overline{\Gamma X}$  (b-d) spin polarization measured at $k_{y}\approx0.05$ \AA$^{-1}$ for (b) $P_x$, (c) $P_y$, and (d) $P_z$. (e) Symmetrized ARPES intensity at $E_F$ in the $k_{x},k_{y}$ plane.}
\end{figure}

\begin{figure*}
\includegraphics[width=1\textwidth]{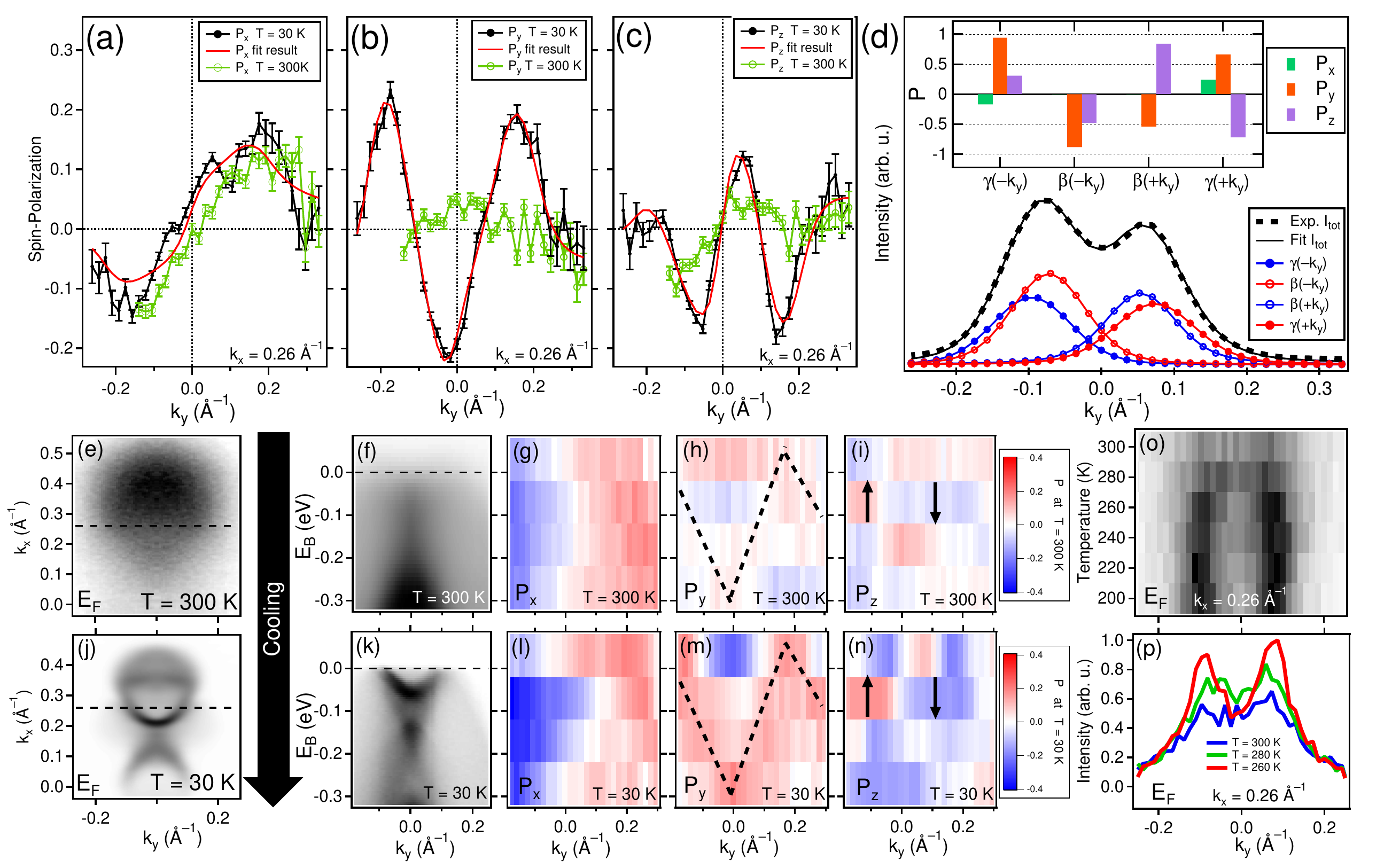} 
\caption{(Color online) (a-c) Spin-polarization momentum distribution curves at $E_F$ and $k_x = 0.26$ \AA$^{-1}$. (d) Results of vectorial spin analysis for the $T = 30$ K data, including peak intensities and spin components (inset). Temperature dependent measurements at (e-i) T = 300 K and (j-n) T = 30 K. (e,j) Fermi surfaces. (f,k) ARPES intensity in gray-scale and spin-polarizations in false-color (see inset) scale for (g,l) $P_x$, (h,m) $P_y$, and (i,n) $P_z$  mapped over $E_{B}(k_{y})$ at $k_x = 0.26$ \AA$^{-1}$. All spin-resolved data were collected using 20 eV photons from the same sample, which was cleaved and measured at 300 K and then cooled. (o,p) Temperature dependence of high-resolution ARPES intensity at $E_F$ and $k_x = 0.26$ \AA$^{-1}$ collected using 67 eV photons for a sample cleaved at 300 K.
}
\end{figure*}

\begin{figure}
\includegraphics[width=0.5\textwidth]{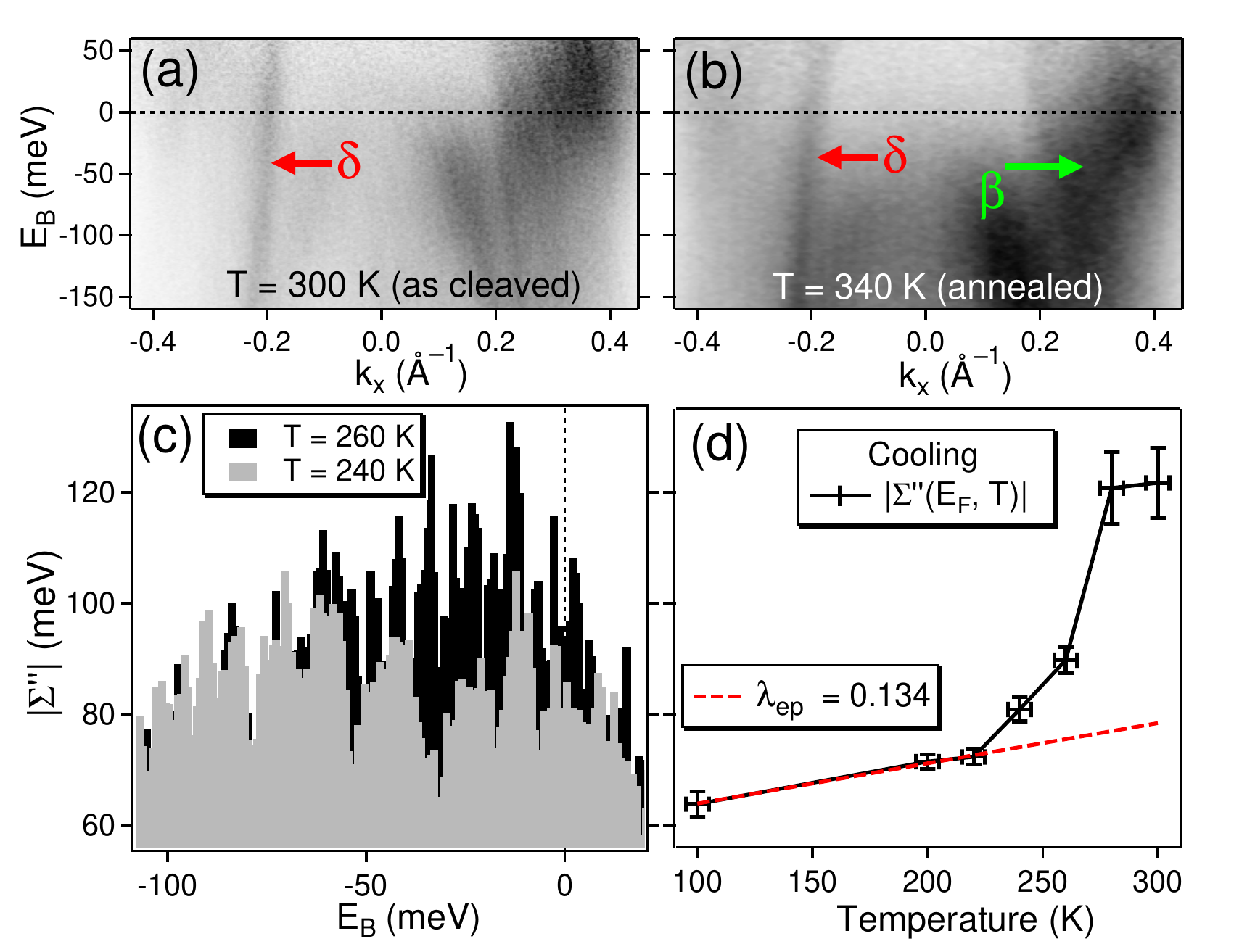} 
\caption{(Color online) (a-b) ARPES intensity along $k_x$ at $k_y$ = 0 divided by the Fermi cutoff. (c) Energy distribution curves of $|\Sigma''|$ in state $\delta$ at $k_{y}=0$ for different temperatures as extracted from raw data. (d) Temperature-dependence of $|\Sigma''(E_{F})|$.}
\end{figure}

Spin-texture visualization provides a resource for understanding scattering amplitudes, spin-transport \cite{Qwang2018}, MR anisotropy \cite{Jiang2015}, and the pairing order and critical field enhancement in superconductivity \cite{Smidman2017}. Spin- and angle-resolved photoemission spectroscopy (SARPES) was used to probe the spin-texture of T$_d$-(Mo/W)Te$_2$ in a few instances \cite{Sakano2017,Feng2016,Crepaldi2017,Jiang2017}, but only small areas of momentum space were covered without measuring the full spin-polarization vector $\mathbf{P}$. Here SARPES measurements and density functional theory (DFT) calculations reveal a spin-texture in the T$_d$-MoTe$_2$ Fermi surface that is three dimensional (3D) both in terms of spin-orientation and momentum dependence. Small and large Fermi arc states persist at more than 90 K above $T_S$. Their coherence improves significantly upon cooling through $T_S$ and the appearance of the large Fermi arc state is affected by thermal history. An anomalous trend of Fermi arc self-energy through the transition and residual spin-polarization in bulk electrons at $T>T_S$ suggest that T$_d$ and 1T' structural phases coexist near the surface at room temperature. A so-called hidden spin-texture \cite{Zhang2014} in 1T'-MoTe$_2$ poses an interesting alternative explanation for the high temperature SARPES results.

Details of the crystal synthesis and DFT calculations are provided in the supplemental material (SM) \cite{Supp}. Experiments were performed with the sample kept under ultrahigh vacuum (UHV) (pressure $<1\times10^{-9}$ Pa) at variable temperatures fully summarized in the SM. Temperature was measured using a Si diode near the sample. Clean (001) surfaces were obtained by cleaving in UHV. High resolution spectra were obtained using a Scienta R4000 analyzer with instrumental angle and energy resolution better than 0.1\textdegree\ and 10 meV. SARPES measurements were done at the COPHEE endstation \cite{Hoesch2002} with angle and energy resolution better than 1.5\textdegree\ and 75 meV. No evidence of mixed (001) and ($00\overline{1}$) terminations \cite{Sakano2017,Tamai2016} was found in our samples \cite{InProg}. ARPES and quasiparticle interference results were consistent with only one termination type \cite{Ruessmann2018}.

Fig. 1 captures the DFT-calculated T$_d$-MoTe$_2$ Fermi surface and its spin-texture, computed as an average over the orbital degree of freedom. The HPs enclosing the $\Gamma$ point of the BZ and EPs located further from the $\Gamma$ point both exhibit high spin-polarization, reaching up to 0.8 in total magnitude \cite{Supp}. This indicates significant orbital anisotropy when compared with, e.g., the Bi$_2$Se$_3$ surface state $(|\mathbf{P}|=0.5)$ \cite{Yazyev2010}. The magnitudes of computed and measured spin-polarization have different significance, because polarized photons selectively excite or entangle orbital components of the electron wave function \cite{Xie2014,Yaji2017,Supp}, but it will be shown that the spin-orientations transform according to crystal symmetries in the same way for both cases. In addition to time-reversal symmetry, the space group contains one reflection $M_x$ and one glide reflection $M_y$ which take the spatial coordinates $(x,y,z)$ to $(-x,y,z)$ and $(x,-y,z+c/2)$, respectively, where $c$ is the unit cell length along the direction perpendicular to the plane of the MoTe$_2$ layers. The in-plane components of $\mathbf{P}$ are constrained by the $M_y$ and $M_x$ symmetries such that $P_x\rightarrow-P_x$ as $k_y\rightarrow-k_y$ and $P_y\rightarrow-P_y$ as $k_x\rightarrow-k_x$, respectively. $P_z$ is constrained to by both of these reflections and must also reverse upon $(k_x,k_y,k_z)\rightarrow(k_x,k_y,-k_z)$. This is a novel property not found in helical spin-textures that suppresses $k_z\rightarrow-k_z$ scattering \cite{Supp}. The spin-polarization remains significant in all components when averaged over the lower half of the BZ as shown in Fig. 1(d-f). This is important for SARPES because the $k_z$-resolution is limited to about $35\%$ of the reciprocal lattice vector due to the finite probing depth.

Fig. 2(a) shows photon energy dependence of ARPES intensity at $E_F$ along $\overline{\Gamma X}$. The states disperse with photon energy, thus characterizing their bulk ($k_z$ dispersive) origin \cite{Xu2016}. The EP around $k_{x}=\pm0.4$ \AA$^{-1}$ (yellow dashed-lines) produces the strongest spin-polarization signal seen in the photon energy dependent maps in Fig. 2(b-d), which are taken along the same direction as in (a), but with a slight misalignment to $k_{y}\approx0.05$ \AA$^{-1}$. This allows $P_x$ and $P_z$, which are reduced to zero by symmetry at $k_{y}=0$, to be measured. The signs of $(P_{x},P_{y},P_{z})$ for the electron states are $(+,+,-)$ for positive $k_x$ and $(+,-,+)$ for negative $k_x$, as enforced by the $M_x$ symmetry. A dissection of the experimental Fermi surface in the ($k_x,k_y$)-plane is shown in Fig. 2(e) in which the contours of bulk EPs (yellow dashed-lines), HPs (blue dashed-lines), large Fermi arc (green dashed-line), and small Fermi arc (red dashed-line) are indicated. The states making up the largest EP, the large Fermi arc, the largest HP, and the small Fermi arc are labelled $\alpha$, $\beta$, $\gamma$, and $\delta$, respectively.

Fig. 3(a-d) shows SARPES measurements of the Fermi surface along the $k_{y}$-direction for $k_{x}=0.26$ \AA$^{-1}$, which crosses through $\beta$ and $\gamma$. The contributions of these states to the spin-polarization shown in Fig. 3(a-c) and intensity in Fig. 3(d) were disentangled quantitatively by vectorial analysis \cite{Meier2008} for the case of $T=30$ K. The momentum distribution curves (MDCs) were fit using four Voight peaks, two for $\beta$ and two for $\gamma$, on a uniform unpolarized background and assuming $\mathbf{|P|}=1$ in each peak. The inset in Fig. 3(d) shows the $P_x$, $P_y$, and $P_z$ values obtained for each peak in green, orange, and purple bars, respectively. The $P_x$ signal primarily originates from hole-like states in this momentum cut, as seen in the binding energy dependence of $P_x$ in Fig. 3(l). The fit results show that $\beta$ and $\gamma$ have opposite signs of $P_y$. In both cases, the sign of $P_y$ is unchanged upon reversal of $k_y$, as required by the combination of $M_x$ and time-reversal symmetry. The sign of $P_z$ in reverses upon $k_y\rightarrow-k_y$ in the case of $\beta$ (which exhibits a negligible $P_x$ component in this momentum cut). $P_x$ and $P_z$ both reverse sign across $k_y=0$ in $\gamma$. Both $\beta$ and $\gamma$ are constrained by \textit{bulk} $M_y$ symmetry, which is broken on the T$_d$-MoTe$_2$(001) surface \cite{Supp}. The quality of the fit with $\mathbf{|P|}=1$ in each state indicates fully coherent spin-orbital coupling at $T=30$ K. Fig. 3(e-p) show measurements taken before and after cooling from 300 K to 30 K. The spin-polarization at $E_F$ for the two temperatures is also compared in Fig. 3(a-c). Response to the temperature change is evident in the lack of a coherent contribution from $\beta$ and overall suppression of $P_y$ and $P_z$ at 300 K. The $P_x$ signal of the hole-like states is retained through the full energy range seen in Fig. 3(g). At both temperatures, hole-like states contribute an M-shape of $+y$-oriented spin in the energy-momentum maps of Fig. 3(h) and Fig. 3(m), as indicated by dashed-lines, and $z$-polarization that switches across $k_y=0$ around $E_{B}=-0.1$ eV, as indicated by arrows in Fig. 3(i) and Fig. 3(n). This serves as a faint signature of T$_d$ order persisting at 300 K.

Spin-integrated MDCs in Fig. 3(o-p) show the development of intensity in $\beta$ upon cooling from 300 K, measured at $E_F$ along the same momentum cut as in Fig. 3(a-c). The peak intensities rise upon cooling from 300 to 280 K, but do not sharpen into clear, Lorentzian shapes until 260 K is reached. One could say that $\beta$ either lies above $E_F$, is fully absent, or the signal is too broad and suppressed to be clearly observed at 300 K. ARPES spectra along $\overline{\Gamma X}$ are shown divided by the Fermi cutoff in Fig 4(a-b). For the case of a fresh surface prepared at 300 K shown in Fig. 4(a), $\beta$ is not visible. It is shown elsewhere that, as in Fig. 3(o-p), $\beta$ does not clearly emerge in this momentum cut either until the sample is cooled to 260 K \cite{Supp}. Fig. 4(b) shows that $\beta$, which presents a line of intensity connecting the bulk electron and hole states (green arrow) \cite{Xu2016,Supp}, persists after cooling to 120 K and annealing to 340 K. It is shown elsewhere that the chemical potential irreversibly increases by about 30 meV upon cooling through $T_S$ \cite{Supp}. It is likely that the spectral function of $\beta$ is broadened and suppressed by scattering in the initial condition, obscuring the signal. These effects may have been diminished by the binding energy shift and/or improved structural order after one thermal cycle. The signal is simply not clear enough in the initial condition for further determination.

The signal of $\delta$, indicated by red arrows in Fig. 4(a-b), is clear at certain emission angles (negative $k_x$) for this case where $p$-polarized 67 eV photons are used. The steep hole-like dispersion reaches above $E_{B}=50$ meV, which is around the maximum energy expected for WPs \cite{Ruessmann2018}. Additional measurements show that $\delta$ corresponds to what ref. \cite{Tamai2016} referred to as a candidate topological surface state \cite{Supp}. To investigate the response of electronic coherence to cooling, the magnitude of the imaginary part of the photohole self-energy $|\Sigma''|$ was computed by multiplying the group velocity with the peak half-width, using raw ARPES data collected at different temperatures \cite{Supp}. There is a significant effect of noise on the results, but it can be appreciated from Fig. 4(c) that there is more area under the distribution of $|\Sigma''(E_B)|$ in the range $-50$ meV $<E_B<0$ meV at 260 K (black bars) than at 240 K (gray bars). Of the possible scattering mechanisms, only electron-phonon coupling (EPC) is expected to cause significant variation in $|\Sigma''(E_{B})|$ near $E_F$ \cite{Valla1999}. In most metals, the lifetime broadening at $E_F$ is proportional to the EPC constant $\lambda_{ep}$ as $|\Sigma''(E_{F},T)|= 2\pi k_B\lambda_{ep}T$, where $k_B$ is the Boltzmann constant \cite{Hof2009}. The average of broadening values extracted from the range $E_F\pm k_B T/2$ is shown as $|\Sigma''(E_{F},T)|$ in Fig. 4(d), with the standard error of the mean shown as error bars. A linear fit in the 220-100 K region obtains a weak dependence on temperature corresponding to $\lambda_{ep}\approx 0.1$ plotted in as a red dashed-line in Fig. 4(d). Linear fitting in the 280-220 K region is unphysical ($|\Sigma''(E_{F},T=0)|<0$). There is a rapid change in EPC, or at least some form of scattering, upon cooling through $T_S$. The strength of EPC has been reported to be similar in 1T' and T$_d$-MoTe$_2$ \cite{Takahashi2017,Heikes2018}, but new forms of electron-lattice interaction arise in the case of strong disorder. For example, electron-phonon-impurity scattering \cite{Reizer1987,Yeh2005}, wherein electron-impurity and electron-phonon scattered paths interfere, can significantly contribute to the self-energy, even at high temperatures \cite{Hsu2012}.

Noting that increased electron density stabilizes the T$_d$ structure \cite{Kim2017,He2018}, it could be that the surface dipole stabilizes local T$_d$ order at temperatures well above $T_S$, in analogy to so-called negative dead layers in ferroelectric materials \cite{Stengel2009}. This would explain the observed residual spin-polarization in bulk electrons. Alternatively, this observation could derive from a so-called R-2 hidden spin-texture \cite{Zhang2014} that must exist in bulk 1T'-MoTe$_2$ because centrosymmetry is absent in all of the lattice sites \cite{Supp}. However, a case of global 1T' order at $T>T_S$ does not explain the anomalous lifetime broadening trend and one would expect a full lattice transition to produce a qualitative change in the measured spin-orientations that is not apparent \cite{Supp}. A mixed structural phase would cause electrons to exist in mixed (non-coherent) states due to entanglement with variations in the lattice-polarity, thus decreasing the quasiparticle lifetimes and spin-polarization magnitudes as resolved by SARPES. 

The results best correspond to a case of local T$_d$ order, with at least one T$_d$/1T' phase boundary existing below the surface at $T>T_S$. There is the added possibility of T$_d$ domains of opposite or unequal lattice-polarity coexisting in this region. Such cases are analogous to ferroelectric polar instability wherein the symmetry-breaking order is short-ranged or fluctuates \cite{Sakai2016}. For MoTe$_2$, this is synonymous with a mixed structural phase due to the first-order nature of the transition, but it has been suggested that the polar instability yields a dynamical or glass-like phase of matter with novel thermoelectric properties \cite{Sakai2016}. The boundary motion is determined by the $c$-axis thermal gradient \cite{Yan2017a}, which is well-defined in the case of a cooled sample with an exposed surface. Boundaries would move into the bulk upon cooling, leaving a globally ordered sample with electronic coherence.

In summary, the observed response to cooling the 1T'-MoTe$_2$ crystal is a gain in electronic coherence that yields a clear view of Fermi arcs and the novel 3D spin texture of T$_d$-MoTe$_2$. The existence of finite $P_z$ must be considered in future discussions of the magnetoresponse properties for T$_d$-(Mo/W)Te$_2$ materials. Both small and large Fermi arc states are observed at 340 K, where the volume of the bulk is almost entirely in the 1T' structural phase \cite{Chen2016nano}. Therefore, the existence of the Fermi arcs is independent of any global, bulk Weyl semimetal phase of matter. Precise determination of the crystal structure near the surface (e.g. by scanning transmission electron microscopy) is vital for clarifying the relationship between the Fermi arcs and the Weyl and structural phases, the anomalous changes in self-energy broadening, and the origin of the spin texture observed at 300 K.

\section*{Acknowledgements}
This work was supported by the Swiss National Science Foundation Project No. PP$00$P$2\_144742$ (no/1), No. 200021-137783, No. PP$00$P2$\_170591$, and NCCR-MARVEL. We thank Titus Neupert and Frank Schindler at the University of Zürich for helpful discussions. P.R., P.M., and S.B. gratefully acknowledge financial support from the DFG (SPP-1666, Project No. MA 4637/3-1) and from the VITI project of the Helmholtz Association as well as computational support from the JARA-HPC Supercomputing Centre at the RWTH Aachen University. 

\footnotesize
\bibliography{References}
\bibliographystyle{apsrev4-1}
 
\expandafter\ifx\csname url\endcsname\relax
 \def\url#1{\texttt{#1}}\fi
\expandafter\ifx\csname urlprefix\endcsname\relax\def\urlprefix{URL }\fi
\providecommand{\bibinfo}[2]{#2}
\providecommand{\eprint}[2][]{\url{#2}}

\end{document}